# Design strategies for optimizing holographic optical tweezers setups

E Martín-Badosa, M Montes-Usategui, A Carnicer, J Andilla, E Pleguezuelos and I Juvells
Grup de Recerca en Òptica Física-GROF, Departament de Física Aplicada i Òptica, Universitat de Barcelona, Martí i Franquès 1, Barcelona 08028, Spain

E-mail: estela.martinb@ub.edu

**Abstract.** We provide a detailed account of the construction of a system of holographic optical tweezers. While a lot of information is available on the design, alignment and calibration of other optical trapping configurations, those based on holography are relatively poorly described. Inclusion of a spatial light modulator in the setup gives rise to particular design trade-offs and constraints, and the system benefits from specific optimization strategies, which we discuss.

## 1. Introduction

The introduction of holographic optical elements into optical tweezer setups has multiplied the possibilities of this technology for precisely trapping, moving and manipulating microparticles. First, static diffractive optical elements generated by computer and manufactured by photolithography, enabled the simultaneous creation of several optical traps [1, 2]. Conversion of the static trap arrays into dynamic light patterns by displaying the diffractive optical elements on spatial light modulators was the logical next step [3-6].

These special displays can be updated at video rates, so that with every new diffractive element a completely different optical potential is formed at the sample plane. Furthermore, holographic optical tweezers have an advantage over other methods of dynamic light array generation, such as time sharing [7], in that the modulator spatially modifies the phase of the incoming wavefronts. Wave-front control easily permits three-dimensional positioning of the traps as well as the creation of beams with special characteristics, such as Bessel or Laguerre-Gaussian beams [8] that carry angular momentum.

In just a few years holographic optical tweezers have become a research topic with many potential applications [9] and thus have turned into a subfield of optical trapping of particular importance and projection. New applications are being proposed in many fields ranging from microfluidics [10, 11] to nanotechnology [12] and biophysics [13]. However, contrary to single-beam technology, which has been thoroughly documented in its many facets [14-18], holographic optical tweezers systems remain comparatively poorly described. The inclusion of the spatial light modulator in the optical setup has important design implications that are specific to this technology. Also, the system may benefit from particular optimization strategies that are worth showing and discussing.

The goal of this paper is to help fill this gap by carefully describing and analyzing the design and construction of a system of holographic optical tweezers, a subject that we have divided into three main parts. Section 1 analyzes the optical system constraints and reveals the trade-offs between optical efficiency and resulting layout size. Several practical tips and design proposals for two different systems are also included here. We believe that the results contained in this section are of wide applicability. Section 2 is devoted exclusively to the spatial light modulator and includes information that is more dependent on the particular device that we use. The three subsections deal with phase-only modulation adjustment, angular dependence of the reconstructed holograms and correction of the built-in optical aberrations of the device. Finally, section 3 contains a complete analysis of the optical aberrations of the resulting optical systems and suggestions on how to achieve diffraction-limited performance with very simple lenses.

**1. Design of the optical setup**
*1.1. Optical system constraints*
Figure 1 shows a diagram of a system of holographic optical tweezers. A continuous-wave, $TEM_{00}$ laser beam is first expanded and then collimated by lenses LE and LC. A pinhole spatially filters the light at the back focal plane of the expander lens, to ensure clean, Gaussian illumination on the spatial light modulator (SLM). The modulator, which may operate either by reflectance or transmittance, is sandwiched between polarizing elements with specific orientations that depend on the configuration desired (see section 2). After passing through a telescope, the beam is reflected upwards by a dichroic mirror and is focused by the high numerical aperture microscope objective on the sample plane. Usually, this latter step takes place inside a commercial inverted microscope, so that the illumination system, objective lens and other optical elements can still be used for imaging purposes. For example, light from the illumination column is transmitted through the dichroic mirror and can reach the sample or the sample plane is still projected on the CCD by the objective and tube lens, allowing observation and recording of the experiments.

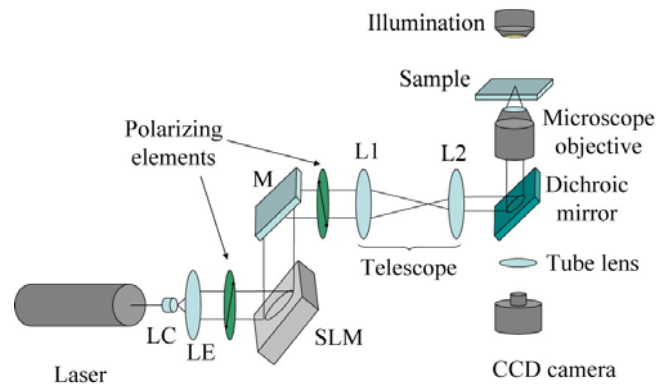

Figure 1. Diagram of the holographic optical tweezers setup.

The telescope formed by lenses L1 and L2 should be designed to meet the following requirements:
1) The SLM is imaged onto the exit pupil of the microscope objective [6, 8] to prevent vignetting of high-frequency Fourier components, which get diffracted at larger angles.
2) To make use of its whole active area, the image of the modulator should be scaled down to match the size of the objective's back aperture. Furthermore, an overfilling of the SLM by the laser will result in an overfilling of the aperture. This, which is necessary, can be accomplished with the first telescope (inverted) formed by lenses LE and LC. The ratio should be adjusted to optimise trapping efficiency as in non-holographic setups [19, 20]. Gaussian laser beams are typically expanded so that the beam waist roughly matches the aperture (here, the size of the SLM).
3) Finally, it must provide parallel illumination to the infinity-corrected objective (Figure 1), hence the telescopic arrangement. Most modern microscope objectives are corrected to work with the sample at the front focal plane. Light rays are therefore parallel after the objective, which is advantageous, since additional optics, such as fluorescence filters or polarizers, can be placed in the path of those parallel rays with negligible effects on focus or aberration correction [21]. An important consequence is that infinity-corrected microscopes need no lenses in the epifluorescence path to collimate light, and thus the two lenses of the telescope can be chosen and placed with total freedom, unlike older microscopes with fixed tube length [18].

Also, regarding the optical system, since the spatial light modulator is illuminated by collimated light and the diffracted beams are observed at the focal plane of the objective lens (focal length, $f$), the relation between the complex reflectance, $R(u,v)$, of the modulator and the electric field at the sample plane, $E(x,y)$, is, except for some often irrelevant phase terms [22], that of a Fourier transform:

$$E(x,y) = e^{i\psi(x,y)} \iint R(u,v) \exp\left[-i\frac{2\pi}{\lambda f}(ux+vy)\right] du\, dv. \tag{1}$$

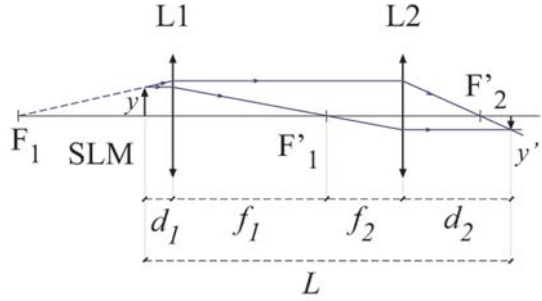

Figure 2. Ray tracing diagram showing the image formation through the telescope L1-L2.

In light of these requirements for the telescope, distance $d_1$ from the SLM to the first lens L1 and distance $d_2$ from the second lens L2 to the objective exit pupil (Figure 2) are subject to several constraints. The distance between these two lenses must be the addition of their focal lengths, $d = f_1 + f_2$, for them to form a telescope. Furthermore, the SLM is imaged by this system with an absolute lateral magnification given by:

$$M = \left|\frac{y'}{y}\right| = \frac{f_2}{f_1}. \tag{2}$$

Incidentally, this magnification is independent of distances $d_1$ and $d_2$, since a ray that leaves the edge of the SLM and travels parallel to the optical axis will always exit the system at the same height (see Figure 2). Total distance, $L$, from the SLM to the objective back aperture is therefore:

$$L = d_1 + d_2 + f_1 + f_2 = d_1\left(1 - M^2\right) + f_1\left(1 + M\right)^2, \tag{3}$$

where equation (2) was considered and distance $d_2$ was calculated by use of the Gaussian lens formula:

$$d_2 = -d_1 M^2 + f_1 M(1+M). \tag{4}$$

The 4-f configuration is a common arrangement for the imaging telescope. The SLM is placed at the front focal plane of the first lens ($d_1 = f_1$) so that the image is formed at the back focal plane of the second lens ($d_2 = f_2$), as shown in Figure 3b. Light rays are parallel between the two lenses and the total length becomes $L = 2f_1(1+M)$. However, this is not the only possibility; Figure 3a and 3c show ray tracings for two alternative arrangements in which the SLM is placed farther away from and closer to lens L1, respectively.

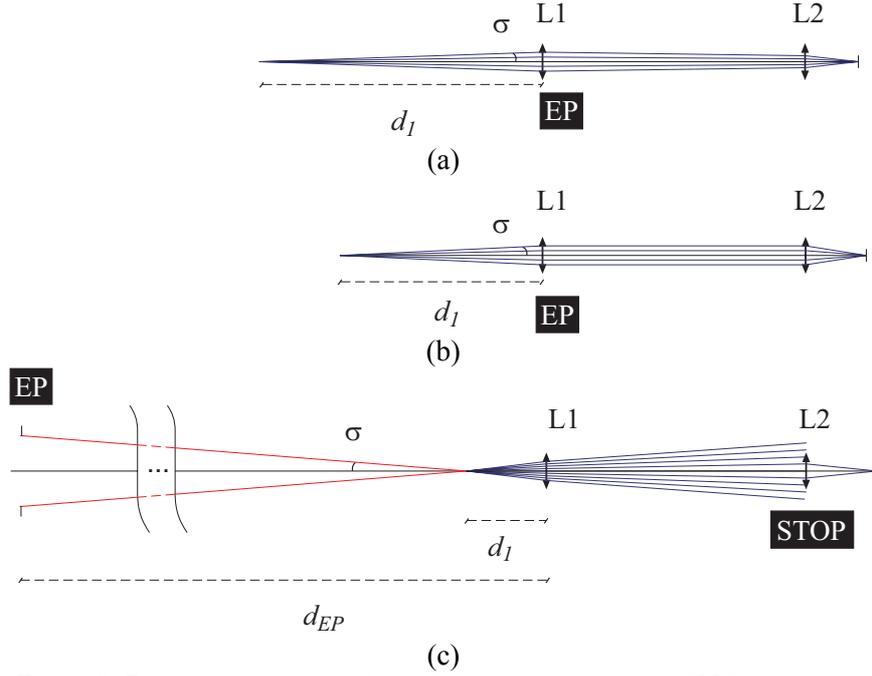

Figure 3. Ray tracing showing the different positions of the SLM with respect to lens L1: (a) $d_1 > f_1$, (b) $d_1 = f_1$ and (c) $d_1 < f_1$. EP, entrance pupil.

Interestingly, more compact setups can be built in this latter case. For a given focal length $f_1$, as magnification $M$ does not depend on SLM position, the shortest overall length $L$ is achieved when the SLM is placed as close to the first lens as possible (minimum $d_1$, equation (3)). The variation of $L$ with $d_1$ for a practical example corresponding to the analysis in section 1.2 ($M = 0.3$, $f_1 = 250$ mm), is depicted in Figure 4a. It can be seen that $L$ approaches its minimum value of $L = f_1(1+M)^2 \approx 420$ mm as the modulator gets closer to lens L1 ($d_1$ tends to zero).

The price for this smaller footprint is lower light efficiency. Indeed, let us assume that the entrance pupil diameter, $\phi_2$, of lens L2 is smaller than or equal to that of lens L1, $\phi_1$, (that is, $\phi_2 \leq \phi_1$). This is normally the case in telescopes, as the diameter of a beam will get smaller at the output. We see that whenever the SLM is placed at a distance $d_1 \geq f_1$ (cases (a) and (b)), lens L1 is both the aperture stop and the entrance pupil of the imaging system. Thus the distance from L1 to the entrance pupil is $d_{EP(a,b)} = 0$ and pupil diameter is $\phi_{EP(a,b)} = \phi_1$. However, when $d_1 < f_1$ (case (c)), L2 acts as the aperture stop of the system and the entrance pupil appears to the left of L1, at a distance:

$$d_{EP(c)} = f_1 \frac{1+M}{M}. \qquad (5)$$

In this case the entrance pupil diameter is $\phi_{EP(c)} = \phi_2 / M$.

The system aperture $\sin \sigma$ is therefore:

$$\sin\sigma = \frac{\phi_{EP}/2}{\sqrt{(\phi_{EP}/2)^2 + (d_{EP}-d_1)^2}} \begin{cases} \sin\sigma_{(a,b)} = \dfrac{\phi_1/2}{\sqrt{(\phi_1/2)^2 + d_1^2}} & (d_1 \geq f_1) \\[2ex] \sin\sigma_{(c)} = \dfrac{\phi_2/M}{\sqrt{(\phi_2/2M)^2 + \left(f_1\dfrac{1+M}{M}-d_1\right)^2}} & (d_1 < f_1) \end{cases} \qquad (6)$$

From equation (6) we can see that for case (c), $\sin\sigma$ increases with $d_1$, whereas in cases (a) and (b) it decreases. Thus, in both situations the maximum value is attained for $d_1 = f_1$. When $\phi_1 = \phi_2$, we have:

$$\sin\sigma_{max}(d_1 = f_1) = \frac{\phi_1/2}{\sqrt{(\phi_1/2)^2 + f_1^2}}. \quad (7)$$

Figure 4b is a graphical representation of equation (6) for $M = 0.3$, $f_1 = 250$ mm and $\phi_1 = \phi_2 = 22.9$ mm. It shows that placement of the SLM at the front focal plane of lens L1 (the 4-f configuration) maximizes light efficiency, although it is intermediate in terms of overall length. The modulator can be placed closer to lens L1 if a smaller optical system is desired. However, as we prefer to retain the light-gathering capacity and shorten the optical system by some other means, as discussed below, a 4-f configuration is assumed for the telescope in what follows.

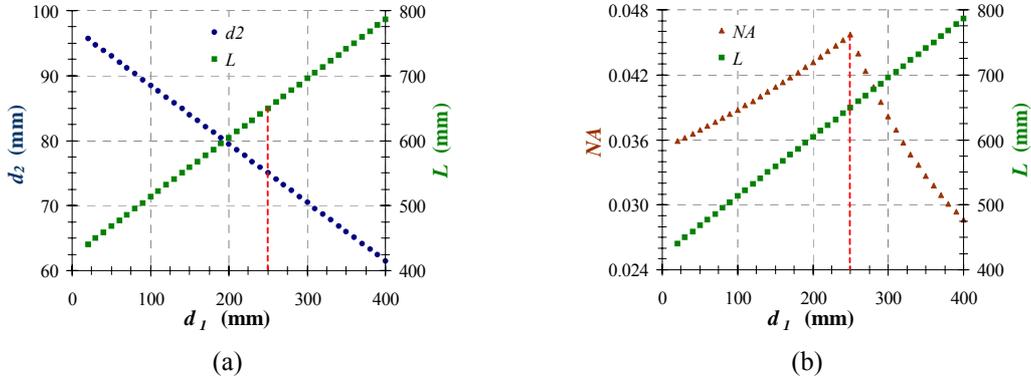

(a) (b)

Figure 4. Numerical example ($M = 0.3$, $f_1 = 250$ mm, $\phi_1 = \phi_2 = 22.9$ mm) of the dependence of $d_2$, $L$ and numerical aperture, $\sin\sigma$, of the telescope with distance $d_1$. The dashed red line indicates the case $d_1 = f_1$.

Let us now focus on the practical constraints for distances $d_1$ and $d_2$. When using an inverted commercial microscope, the minimum distance from lens L2 to the exit pupil of the objective, $d_2$, is some 300 mm, if the lens is placed outside the microscope (roughly the length of the fluorescence path). This limitation determines to a great extent the overall size and often leads to large optical systems: in effect, for $d_1 = f_1$, $d_2 = f_2$ (4-f arrangement), equation (3) becomes:

$$L = 2(f_1 + f_2) = 2d_2 \frac{1+M}{M} \quad (8)$$

The active area of the spatial light modulators used in optical tweezers range from about 8 mm (BNS P512 [23]) to about 20 mm (Hamamatsu X8267 [24]) on a side. However, the exit pupil diameter of high-aperture, immersion objectives is between 3 and 5 mm. Thus, lateral magnification $M$ takes values between 0.1 and 0.6. For a typical value of $M = 0.4$ and with $d_2 = 300$ mm, $d_1$ becomes 750 mm and finally $L = 2.1$ m.

Such long working lengths help to minimize optical aberrations [2] and are thus frequently viewed as a desirable feature. However, our results indicate (section 3) that aberrations in the optical train are not really an issue and can be controlled quite easily by use of a few simple tricks. Thus, distance $L$ could be much shortened if $d_2$ were reduced by placing lens L2 inside the microscope. Practical details on how to do this are left to the proposed solution in the next section 1.2.

Reduction in distance $d_2$ leads to reduction in distance $d_1$, as both are linked by equation (2). This latter distance is subject to design constraints of its own and frequently cannot be made smaller than a certain limit, which should be taken into account when reducing the design.

For example, minimal distance between SLM and lens L1, $d_1$, can be just a few centimetres for transmittance SLMs [3, 4], the space required to allow polarizing elements to fit between. However, reflective SLMs are more commonly used as they provide higher resolutions and a larger fill factor. They do pose different geometrical requirements. One possible arrangement would be to place the reflective modulator perpendicular to the optical axis [25] and then redirect the modulated beam out with a beam-splitter, as shown in Figure 5a. However, control of the input and output polarization is a much desired feature of the setup, as it may allow free access to the different operating modes of the device (such as the phase-only modulation operating curve). Both constraints, polarization control and on-axis operation, can be met by the use of a non-polarizing beam-splitter, but the round-trip path through that element would result in a loss of 75% of incident light. This is unacceptable given the large power required for trapping even a small number of samples, so the device is usually operated at small incident angles $\theta$ (Figure 5b) [26]. This sets the minimal distance at:

$$d_{1\min} = \frac{\phi_{SLM}}{2\tan\theta}, \qquad (9)$$

the distance required for separating input and output beams.

Figure 5c shows distance $d_{1min}$ as a function of $\theta$ for a medium size SLM, $\phi_{SLM} \sim 14.6$ mm (Holoeye Photonics, LC-R 2500, see section 2, below). For small angles, $\theta \sim 2.5°$, minimum distance is $d_{1min} \sim 170$ mm. Allowance for polarizing elements would add some centimetres to the total count. At any rate, in practice, this result appears to be less restrictive than limitations arising from the position of lens L2.

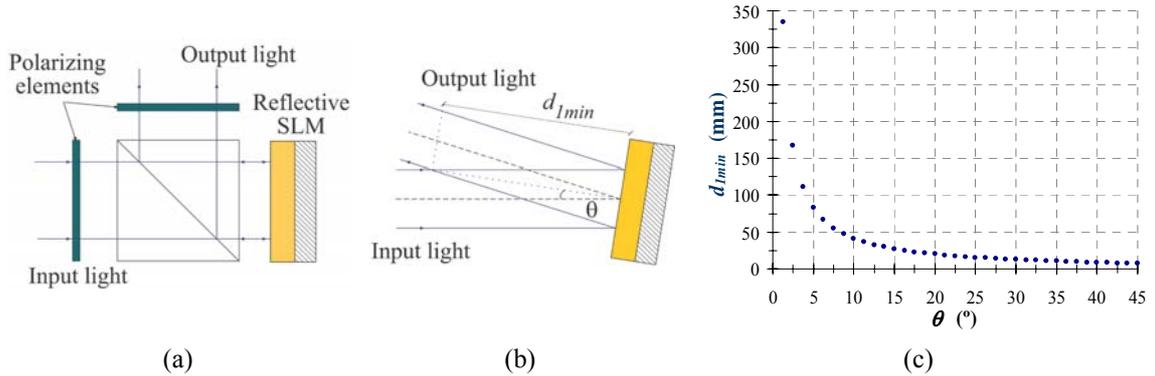

(a)            (b)            (c)

Figure 5. Geometry in a reflective SLM: (a) with a beam-splitter and (b) by tilting the SLM. (c) Dependence of the minimal distance $d_{1min}$ on the incident angle $\theta$.

*1.2. Optical setup: proposals and practical considerations.*

According to the analysis of the optical system constraints carried out in the preceding section, we built a system of holographic optical tweezers as follows: a Nd:YVO$_4$ laser beam (Viasho Technology, $\lambda = 532$ nm, 120 mW) illuminates a twisted-nematic liquid-crystal reflective SLM (Holoeye Photonics, LC-R 2500), sandwiched between a polarizer and an analyzer with the proper orientations to achieve phase-mostly modulation (see section 2). Light after the telescope enters a commercial inverted microscope. We show results for two different models, a Nikon Eclipse TE-2000E and a Motic AE-31, equipped with oil-immersion, high-numerical aperture objectives (Nikon Plan Fluor 100x, 1.30 NA and Motic Plan achromatic 100x, 1.25 NA, respectively).

Distance $d_2$ from lens L2 to the objective exit pupil is greatly reduced by placing the lens inside the microscope body. This is straightforward for the TE-2000 microscope, as a lens can be easily retrofitted into standard Nikon fluorescence cubes. Each cube has 1'' threaded circular aperture on one of its six surfaces, for the excitation filter to be mounted. However, lenses or any other optical or mechanical component can be attached there instead. Figure 6a shows a photograph of a cube with the dichroic mirror inside and lens L2 on the input face, directly screwed into place. Although not visible, an absorbance filter is also present on the bottom side (exit) to filter the laser out, thus preventing camera saturation.

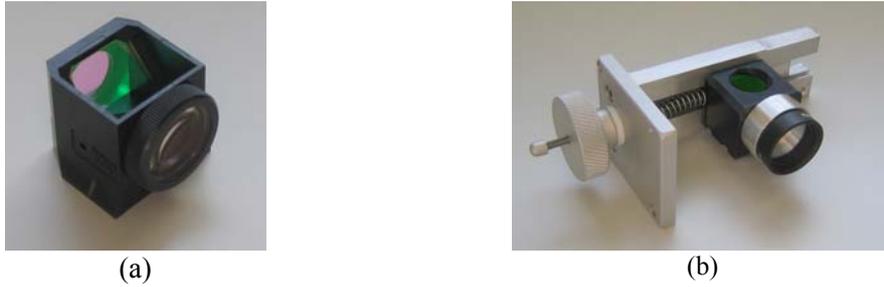

(a)           (b)

Figure 6. Detailed photographs of (a) a standard commercial cube for a Nikon Eclipse TE-2000E microscope and (b) a custom-made cube for a Motic AE-31 microscope.

However, custom-made modification was necessary for the Motic AE-31 cube, an inexpensive inverted microscope with fluorescence capacity. A centring mechanical component containing the cube is shown in Figure 6b.

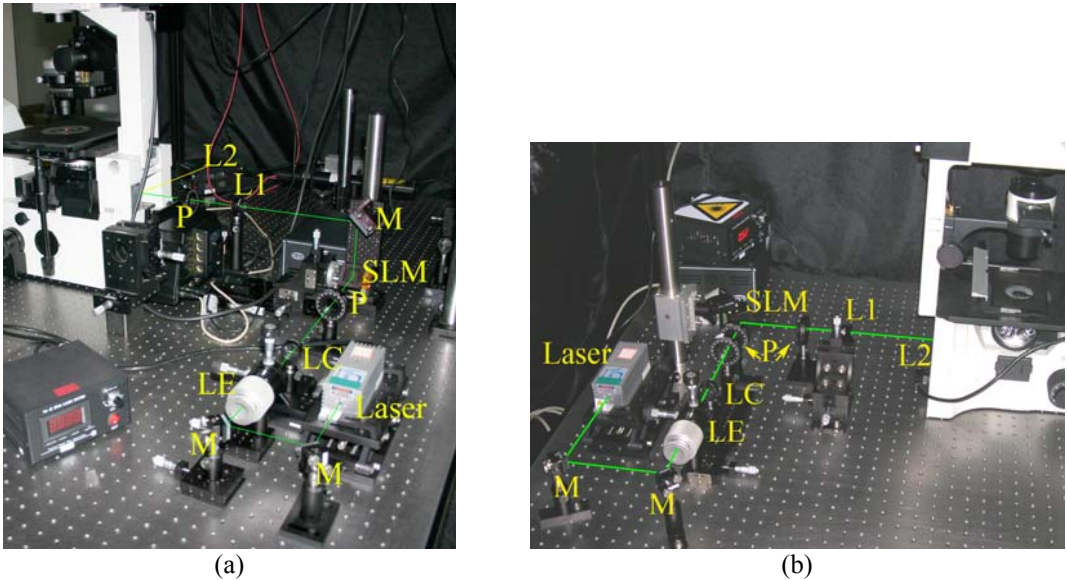

(a)           (b)

Figure 7. Photographs of two holographic optical setups with microscopes (a) Nikon Eclipse TE-2000E, (b) Motic AE-31. (M: mirror, P: polarizing element)

By adopting this solution, distance $d_2$ goes down to about 75 mm for the Nikon microscope and to 100 mm for the Motic instrument. If the vertical size of the Holoeye modulator (see specifications in section 2) is imaged by the telescope to match the objective exit pupil (diameter of about 4 mm, Nikon Plan Fluor 100x), the magnification would then be $M = 4/14.5$. If we place the SLM on the front focal plane of lens L1 then $d_1 \approx 275$ mm (corresponding to $d_2 = 75$ mm), the total length between the SLM and the microscope objective becomes $L \approx 0.7$ m. This is much shorter than the 2.1 m we obtained in section 1.1, especially as part of the path is inside the microscope. Conveniently, the distance between lenses L1 and L2, which is about $75 + 275$ mm $= 350$ mm, is large enough to enable lens L1 to be placed outside the microscope. Similar conclusions can be drawn for $d_2 = 100$ mm (Motic microscope). Finally, the required tilt $\theta$ of the SLM ($\phi_{SLM} \approx 15$ mm) with respect to the incident beam (Figure 3b), such that it can be put as close to lens L1 as $d_1 \approx 275$ mm, is only about a few degrees, according to equation (9). Nevertheless, we place the SLM tilted 45º, because it simplifies the arrangement of the whole optical setup. We found that, although not lying on a plane perpendicular to the optical axis, the SLM is capable of producing fairly good traps when functioning in a suitable phase-only configuration. These issues are analyzed in detail in the next section.

Figure 7 shows pictures of the optical setups built for the Nikon Eclipse TE-2000E (Figure 7a) and for the Motic AE-31 (Figure 7b) microscopes. We see the expander lens LE, collimating lens LC and some polarizing elements before and after the SLM, which is arranged to reflect light at right angles. Placing the SLM at 45º considerably simplifies the optical setup, as it works geometrically as a mirror. Both lenses L1 are outside the microscope bodies and lenses L2, mounted on their respective cubes, cannot be seen.

## 2. Holoeye LC-R 2500 spatial light modulator

The Holoeye LC-R 2500 used in our setups is a reflective, twisted nematic, liquid crystal on silicon (LCOS) spatial light modulator (SLM), and was selected for its high resolution, good optical quality and low cost. It has an active area of 19.5x14.6 mm$^2$, divided into 1024x768 square pixels (pixel pitch = 19 μm, fill factor 93%) that are electrically addressed by an 8-bit signal coming from a computer graphic card through the DVI interface [27]. The SLM can provide from 0 to $2\pi$ phase modulation in the visible region (400-700 nm), although usually with a certain amount of amplitude modulation, which is inherent to twisted nematic liquid crystal modulators [28]. The response depends on the polarization state of the light both before and after the SLM and can be changed by adjusting the polarization elements shown in Figure 7. Since we are not using the device at the near-normal incidence assumed by the manufacturer [26], modulation response was measured and adjusted.

In section 2.1 we give details on these measurements and report a configuration for which there is nearly uniform amplitude modulation and a maximum phase shift of $2\pi$ (phase-mostly configuration), when the device makes an angle of 45º with the incident light. Also, in section 2.2 we analyze the effect of oblique incidence on the reconstruction of holograms. Finally, section 2.3 is devoted to the correction of certain aberrations of the display caused by the curvature of the silicon backplane.

### 2.1. Characterization: phase-only configuration

Spatial light modulators are electro-optic devices that can modify (modulate) the amplitude and phase of an optical wavefront on a controlled basis. Spatial light modulators come in many flavours [29], the most frequently employed for holography being those based on liquid crystal mixtures. Modulation is achieved by making light propagate inside an optically active medium (the liquid crystal), whose optical properties, notably extraordinary index of refraction or optical axis, are voltage-dependent. Thus, the retardance induced by such a device is, in turn, dependent on the voltage applied to the liquid crystal cell, as well as on the polarization of the traversing light beam.

Spatial light modulator characterization consists of determining the amplitude and phase modulation values for the different addressing voltages, for a fixed orientation of both input and output polarization elements. For an 8-bit modulator, up to 256 different voltage values can be applied to any given pixel. Usually, SLM electronics is connected to the graphics board of a computer and the user feeds the SLM by displaying images on the PC screen. The different grey levels ($gl$) of the image (ranging between $gl = 0$ and $gl = 255$) are eventually translated into driving voltages inside the modulator.

For each grey level, amplitude modulation is simply measured as the square-root of the transmittance of the device. Phase modulation is the phase change imparted by the modulator onto the incoming wave front. Since only phase differences are of any physical significance, it is obtained by measuring the relative phase shift between a reference grey level (phase modulation arbitrarily set to zero) and the other grey-level values.

Figure 8a shows a sketch of the Mach-Zehnder interferometer that we used to measure the phase shifts: a collimated laser beam ($\lambda = 532$ nm) is divided by the first beam-splitter and a CCD camera records the interference fringes between the two plane waves. The orientations of the quarter-wave plate and the two polarizers determine the modulation properties of the SLM, which is placed in one arm of the interferometer. The phase it introduces is observed as a fringe displacement on the interference plane. When a constant reference grey level is displayed on one half of the modulator and the remaining values are sequentially displayed on the other half, shifted interference fringes can be seen on the CCD image. The relative phase shift $\varphi$ can be accurately measured with the fringe analysis method described in [30], as:

$$\varphi = 2\pi \Delta / P, \qquad (10)$$

where $P$ is the period of the fringes and $\Delta$ is the relative fringe displacement. The former is determined by Fourier transforming the fringe images, while the latter is obtained by performing the one-dimensional correlation product between the shifted fringe patterns. This gives a measurement of the similarity between the two functions as one of them is displaced over the other. The position of the correlation maximum corresponds to the maximum overlapping conditions and, in the case of fringe correlation, it directly provides the fringe displacement.

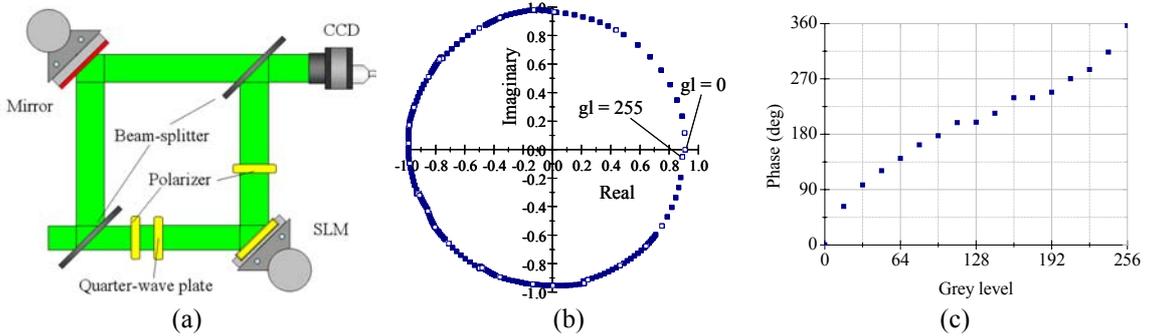

(a) (b) (c)
Figure 8. (a) Mach-Zehnder interferometer for SLM phase determination. Holoeye LC-R 2500 phase-mostly configuration at 45º (b) representation in the complex plane and (c) phase as a function of the grey level (gl).

Finally, the setup in Figure 8a also allows transmittance measurements, by placing an intensity detector before and after the SLM for each displayed grey-level value; amplitude modulation is simply the square-root of the transmittance, and is usually normalized to one.

The best phase-mostly operating curve we found is shown in Figure 8b. The graph is a polar plot where the amplitude and phase modulation are jointly displayed as a complex-plane curve, each point standing for a single grey level (from $gl = 0$ to $gl = 255$). The magnitude of a vector from the origin of coordinates to the point considered (i.e., the radial coordinate) gives the amplitude modulation; and the angle with respect to the real-positive axis (the polar angle) is the phase shift. The measurement was done for 16 evenly spaced grey-level values, represented by empty squares in the curves. The remaining points (solid squares) are obtained by linear interpolation. This configuration is achieved when the modulator is sandwiched between two linear polarizers, oriented at -45º and 26º with respect to the longest side of the display. Positive angles are measured counter-clockwise, when looking at the polarizer by the side first touched by the laser. Maximum phase shift reaches $1.98\pi$, amplitude is almost constant with an intensity contrast of only 1:1.25, and optical efficiency is around 50%. Finally, Figure 8c shows the phase modulation as a function of the grey level (addressing voltage), which is not uniform and should be linearized when the holograms are computed.

*2.2. Reconstruction of holograms by a tilted modulator*
As explained above, we place the modulator at 45º with respect to the beam direction. Here, we analyze how this may change the reconstruction of the displayed hologram $R(u,v)$, and affect the performance of the optical traps.
The hologram $R(u,v)$ is calculated by means of equation (1). Usually, the spatially-variant phase term $e^{i\psi(x,y)}$ is not taken into account as it disappears when the intensity of the diffracted light is recorded. Our goal in this section is to study how the hologram is reconstructed when the modulator is tilted, which means that the distance between each SLM row and the objective changes (Figure 9a). As $e^{i\psi(x,y)}$ explicitly depends on the distance between the hologram and the front principal plane of the objective, we need to include this phase in the analysis. The complete formula that links a hologram $R(u,v)$ with the reconstructed field $E(x,y)$ [22] is:

$$E(x,y) = \exp\left(\frac{ik}{2f'}\left(1-\frac{d}{f'}\right)(x^2+y^2)\right) FT_{\lambda f'}[R(u,v)], \qquad (11)$$

where $d$ is the distance between the SLM and the front principal plane of the objective, $f'$ is the effective focal length ($f' = f/n$, where $n$ is the index of refraction), and $FT_{\lambda f'}[R(u,v)]$ stands for the Fourier transform of $R(u,v)$ evaluated at frequencies $x/\lambda f'$ and $y/\lambda f'$. Notice that $|E(x,y)|^2 = |FT_{\lambda f'}[R(u,v)]|^2$.

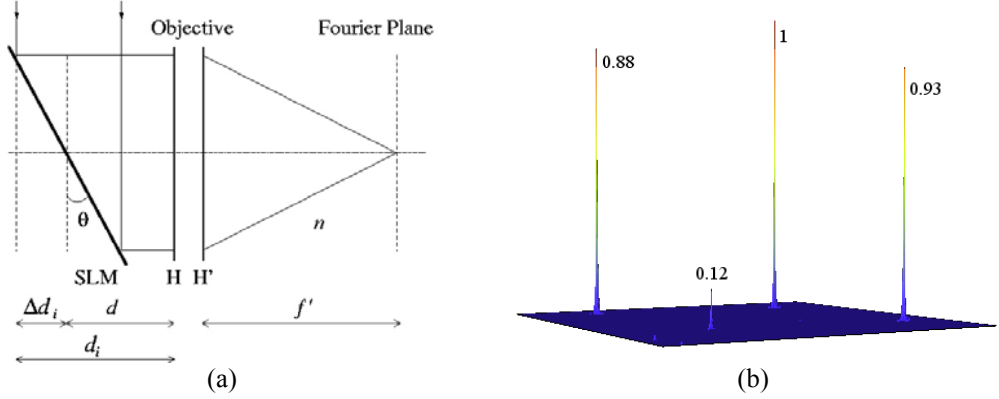

(a) (b)

Figure 9. (a) SLM titled by an angle $\theta$ with respect to the beam direction ($n$: index of refraction). (b) Simulation of the diffraction effects arising from equation (14) ($\theta = 45°$) on three equal intensity traps.

The modulator is tilted in such a way that the distance $d_i$ between the $i$th modulator row and the front principal plane $H$ is constant, but varies from row to row. Thus, $d_i = d + \Delta d_i$. Then, the contribution of the $i$th row $R(u,v_i)$ to the total electric field is:

$$E_i(x,y) = \exp\left(\frac{ik}{2f'}\left(1-\frac{d_i}{f'}\right)(x^2+y^2)\right) FT_{\lambda f'}[R(u,v_i)], \qquad (12)$$

where $FT_{\lambda f'}[R(u,v_i)]$ is the two-dimensional Fourier transform of row $R(u,v_i)$. Therefore, the proper expression for the reconstructed field can be written by addition of terms similar to that in equation (12), for different distances $d_i$.

$$E = \exp\left(\frac{ik}{2f'}\left(1-\frac{d}{f'}\right)(x^2+y^2)\right) \sum_i \exp\left(-\frac{ik}{2f'}\frac{\Delta d_i}{f'}(x^2+y^2)\right) FT_{\lambda f'}[R(u,v_i)]. \qquad (13)$$

Finally, the intensity is

$$I \propto EE^* = \left|\sum_i \exp\left(-\frac{ik}{2f'}\frac{\Delta d_i}{f'}(x^2+y^2)\right) FT_{\lambda f'}[R(u,v_i)]\right|^2. \qquad (14)$$

Notice that this result is independent of distance $d$. An example simulated using this equation is shown in Figure 9b. A hologram was computed by using the Gerchberg-Saxton algorithm [31, 32] to obtain three equal intensity traps. The simulation realistically takes into account both the actual phase-mostly modulation curve and the geometrical parameters of the SLM (section 2). The figure shows a typical result: as a consequence of the SLM tilt, the intensity of the three traps is no longer uniform (a small replica is also visible). These differences are hardly noticeable experimentally and can be corrected during hologram computation if desired. Our conclusion is that use of the SLM at large angles to the optical axis does not seem to pose any major difficulty and may simplify the optical layout. A different but related study that supports similar conclusions can be found in reference [33].

*2.3. Modulator aberrations and correction*

A liquid crystal on silicon (LCOS) micro-display essentially consists of a liquid crystal layer between a cover glass and a silicon backplane that contains the driving circuitry and is coated with aluminium to form a highly reflective surface. The backplane is manufactured at commercial VLSI foundries using standard CMOS methods and processes, which unfortunately are not optimized with optical performance in mind. In consequence, although the resulting devices can be made flat at the pixel level to prevent scattering and improve diffraction efficiency, they have bows and warps at the die level [34], which frequently gives rise to major optical aberrations.

In particular, the beam reflected by our Holoeye modulator, when focused by the microscope objective, forms two light lines instead of a diffraction-limited spot. The two lines are perpendicular to each other and appear at two different focal planes in a behaviour similar to the presence of astigmatism (refer to Figure 10). Figure 10b (left and right) corresponds to the images captured at the two foci when a pair of optical traps is generated. The effect, which also takes place at normal incidence, reveals the lack of flatness of the device surface and seems to be a widespread problem [35].

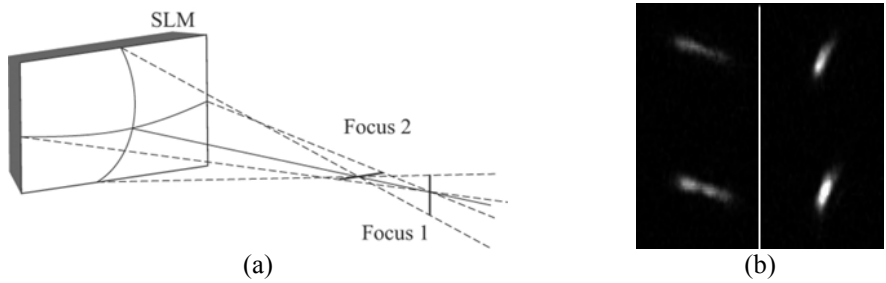

(a)  (b)
Figure 10. (a) Light reflected by the SLM converging at two different foci.
(b) Intensity of a pair of optical traps at the two foci.

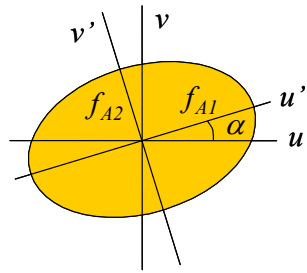

Figure 11. Definition of the parameters used to correct SLM aberration.

Fortunately, in our case it was relatively easy to find an anamorphic phase function that reverses the effect and which can then be added to any trapping hologram, eventually correcting the aberration. The correction is modelled after the following phase function, $\phi_{ab}$:

$$\phi_{ab} = -\frac{\pi}{\lambda}\left[\left(\frac{u'}{f_{A1}}\right)^2 + \left(\frac{v'}{f_{A2}}\right)^2\right], \tag{15}$$

which is the quadratic approximation of an elliptical wave having focal lengths $f_{A1}$ and $f_{A2}$ in the $u'$ and $v'$ directions, respectively [22]. Here, $u'$ and $v'$ are the $\alpha$-rotated axes of $u$ and $v$ (the modulator horizontal and vertical coordinates, Figure 11):

$$u' = u\cos\alpha + v\sin\alpha$$
$$v' = v\cos\alpha - u\sin\alpha \qquad (16)$$

We found experimentally by trial and error parameters $f_{A1}$, $f_{A2}$ and $\alpha$ which satisfactorily correct the aberration. We display the corresponding phase function $\phi_{ab}$ on the SLM and observe the quality of the optical trap: initial values for $f_{A1}$ and $f_{A2}$ are determined by finding the two foci in Figure 10a, then different phases are computed with various $\alpha$ orientations (while allowing slight changes in $f_{A1}$ and $f_{A2}$) until a trap with circular symmetry is obtained. We finally found $f_{A1} = 30$ m, $f_{A2} = 8$ m and $\alpha = 17°$, for our modulator. The resulting phase correction, adapted to the SLM operating curve (figures 8b and 8c), is shown in Figure 12a. Figure 12b shows the captured images of the two traps at the mid-point of the two foci (left) and the spots after correction (right). A more quantitative approach to characterizing and correcting SLM aberrations can be found in [35]. A final possibility for determining phase correction would be to use a Shack-Hartmann wave-front sensor, as this allows accurate and automated measurements of the wave aberration.

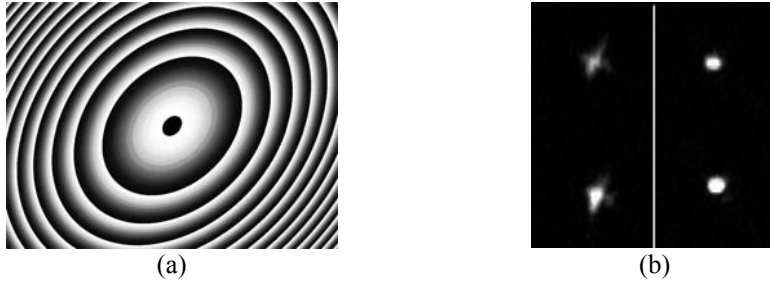

(a)                     (b)

Figure 12. (a) Elliptical phase that corrects the aberration of Holoeye modulator. (b) Image of two optical traps without correction (left) and after correction (right).

### 3. Aberrations in the optical train

Single-beam optical tweezers rely on the gradient component of the light force to trap and move material particles. As light has to be brought to a sharp focus, exquisitely corrected microscope objectives are required to form the optical traps. Optical aberrations are always a major cause of concern, since their detrimental effect on trap stiffness is well documented [36]. Because of this, an accurate analysis of aberrations of the complete optical system should be included in the design process as a final stage.

In this section, we perform ray-tracing simulations with commercial optical design software (Zemax) to study the spherical aberration in the focal plane of the microscope objective, for different qualities and combinations of outside lenses. The microscope objective is considered a paraxial lens to isolate the influence of the other lenses on the quality of the focal spot. This approach may provide useful rules of thumb for the design of the external setup but it should be noted that important aberrations may be left out of the analysis. For example, when using oil-immersion lenses, the index mismatch at the glass/water interface is known to be a major source of spherical aberration [37].

As our objectives are infinity-corrected, spherical aberration is minimized at the front focal plane. The focal length in air, $f$, can be obtained from the magnification, $m$, and the focal length of the microscope tube lens, $f_{tube}$, $f = f_{tube}/m$ [21]. Nikon and Motic objective focal lengths are 2 mm and 1.2 mm, respectively.

This section is organized as follows: first, in section 3.1 a perfectly parallel beam is supposed to illuminate lens L1 to isolate the effect of the imaging telescope formed by lenses L1 and L2. In section 3.2, the best solution found in section 3.1 is analyzed jointly with the beam expander and collimator. Whenever possible, we use commercial standard single lenses. Unless otherwise specified, these are made of BK7, which has good transmission throughout the visible and the near infrared spectra (over 90%). Also, in practice, antireflection coatings should be used to help reduce transmission losses and stray reflections.

Aberrations introduced by the SLM are not taken into account, as they have presumably been corrected with the method developed in section 2.3. However, we do consider that the modulator is tilted by 45º in the horizontal (long) direction and that the illuminated area of the device, as well as the active area imaged onto the exit pupil of the objective, has an elliptical shape.

### 3.1. Imaging telescope (L1 and L2)

Again, we assume a 4-f configuration for the telescope (section 1) and we analyze the setup for the Nikon microscope, which needs shorter focal lengths than the Motic and is therefore more prone to aberrations. Required lens L2 focal length is $f_2 = d_2 = 75$ mm and lens L1 focal length is chosen as $f_1 = 250$ mm ($M = 0.3$). This is a standard value in many vendor catalogs, making it an easy lens to find commercially, and one which fully meets magnification requirements: for an exit pupil diameter of 4 mm (Nikon Plan Fluor 100x objective), the SLM active area imaged onto the objective exit pupil is elliptical, with its vertical and horizontal axis of about 700 and 1000 pixels, respectively. The total length given by equation (8) is in this case $L = 650$ mm.

We chose plano-convex lenses, since these are the best singlet form for either focusing collimated light or for collimating a point source. Thus, placing two plano-convex lenses with their flat surfaces facing one another is the best simple solution of spherical aberration for a telescope [38]. Figure 13a shows the spot diagram obtained in the focal plane of the microscope objective with this configuration. The results are worse if the same plano-convex lenses are used but with a wrong orientation (curved surfaces facing one another, Figure 13b) or for two bi-convex lenses (results not shown). Inexpensive standard Thorlabs 1'' BK7 singlet lenses are used in our simulations: LA1461 (250 mm plano-convex), LA1608 (75 mm plano-convex), LB1056 (250 mm bi-convex) and LB1901 (75 mm bi-convex).

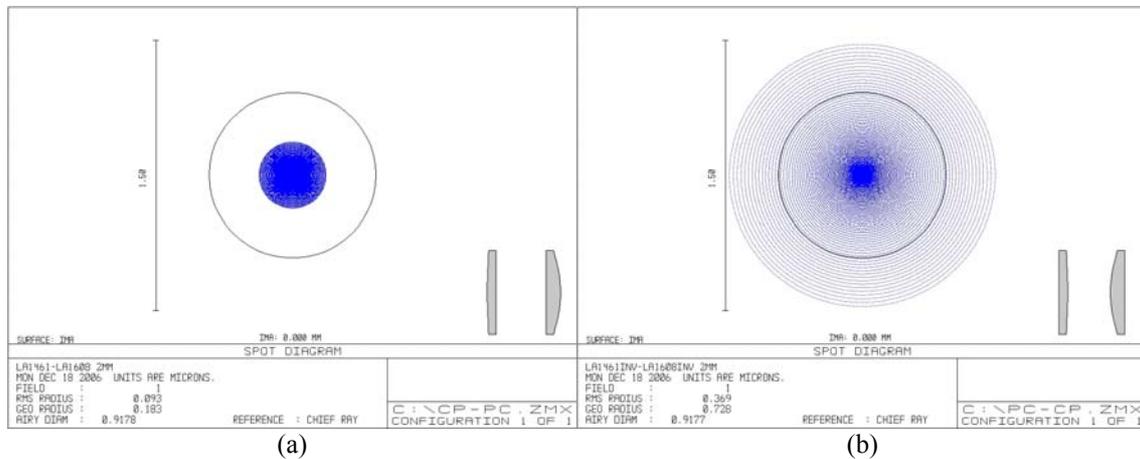

(a) (b)

Figure 13. Spot diagrams for the following shapes and orientation of lenses L1 and L2: (a) plano-convex, curved surfaces facing collimated beams; (b) plano-convex, flat surfaces facing collimated beams.

The RMS radius (root-mean-square radial size) indicated on the plots gives an approximate idea of the spread of the ray bundle. Roughly speaking, if all rays are well within the Airy disk (represented in the figures by a circle), the system is often considered to be diffraction-limited.

The results show that two plano-convex lenses with their curved surfaces facing the collimated beams (Figure 13a) produce a diffraction-limited spot (RMS radius = 0.093 μm, Airy radius = 0.46 μm), whereas in the opposite case, shown in Figure 13b, they do not (RMS radius = 0.369 μm). Also, if the telescope is made with two bi-convex lenses, the result is intermediate (RMS radius = 0.137 μm).

In conclusion, simple plano-convex lenses are a good enough choice for the imaging telescope. There is no need for more sophisticated optics, even in systems like the one we are analyzing that reduce dimensions by use of lenses with short focal lengths (such as $f_1 = 250$ mm, $f_2 = 75$ mm).

*3.2. Beam expander and collimator (LE and LC)*
As mentioned in section 1.1, the laser beam expander and collimator also form a telescope, although used in reverse. It will increase the beam diameter of an incident Gaussian beam by:

$$\tilde{M} = \frac{f_C}{f_E} \qquad (17)$$

while simultaneously reducing the beam angular divergence by the inverse factor $1/\tilde{M}$. We chose for our simulations $f_E = 8$ mm, $f_C = 100$ mm so that a laser beam diameter of about 1.1 mm (like ours) is magnified 12.5 times. The beam then illuminates some 720 vertical pixels of the Holoeye modulator, fully consistent with the imaged active area we set in section 3.1 and overfilling requirements.

We assume that lens LE is a paraxial lens and we focus our analysis on collimating lens LC alone. This is because LE is either not necessary, when a fibre laser is used, or is a well-corrected microscope objective, when light is expanded by a spatial filter. In any case, the collimator receives a high-quality Gaussian beam that we assume is free from aberrations.

Figure 14a shows the results for an inexpensive plano-convex singlet (Thorlabs LA1509, 1'', BK7, 100 mm) with the right orientation. Under these conditions, the spot is not diffraction-limited (RMS radius = 1.208 μm, Airy radius = 0.46 μm), which is not surprising. For example, when compared to lens L2, even though its focal length is somewhat larger ($f_C = 100$ mm, $f_2 = 75$ mm), it works at a higher aperture (related by magnification $M$). Lens LC is the most critical element in the system with regard to aberrations and needs higher-quality optics.

We tried an achromatic doublet corrected in the visible region: 1'' Thorlabs' lens made of BK7 and SF5 (item number AC-254-100-A1). The result, plotted in Figure 14b (RMS radius = 0.244 μm), shows that spherical aberration is substantially reduced and that the spot can now be considered diffraction-limited.

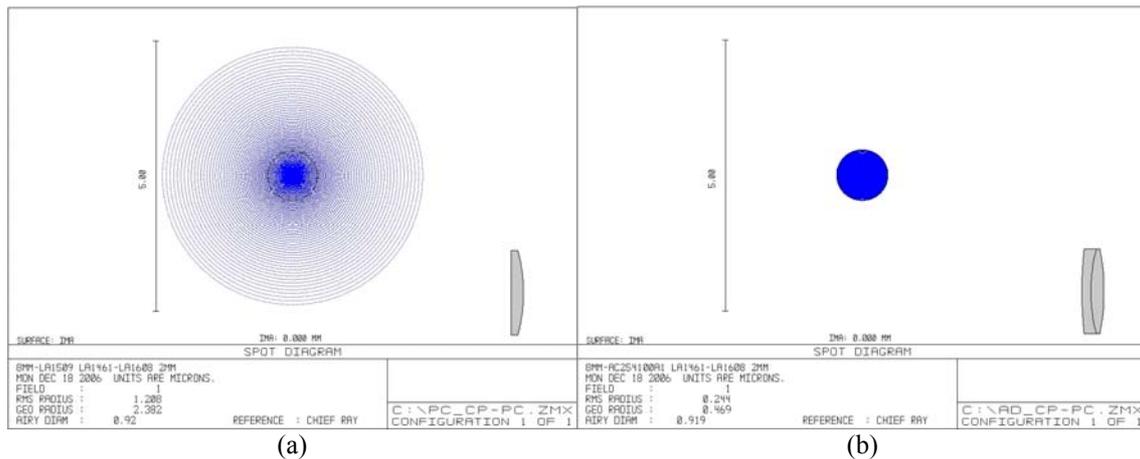

(a) (b)
Figure 14. Spot diagrams when the following shape of lens LC is used: (a) plano-convex, (b) achromatic doublet.

In conclusion, special care should be taken when selecting lens LC. Because of its working conditions, it is especially prone to spherical aberration; and so a corrected lens, such as a doublet or an aspheric one, needs to be used.

## 4. Results and final remarks

Figure 15 is illustrative of the characteristic results that may be obtained with the system. Many quality traps showing diffraction rings can be simultaneously produced. Importantly, the central spot is very small (see Figure 15a), which we believe is indicative of a well-controlled spatial light modulator. A large DC term creates an unwanted trap at the centre of the sample and may need to be filtered out. A deficient phase-only curve or failure to correct the non-linearities between driving voltages and phase values (Figure 8c) in computing the holograms contributes to this central spot. Light reflected at the front face of the SLM or unmodulated because of fill factor issues are also small contributions that cannot be eliminated.

Finally, figures 15b and 15c show polystyrene microspheres trapped and moved independently in three dimensions.

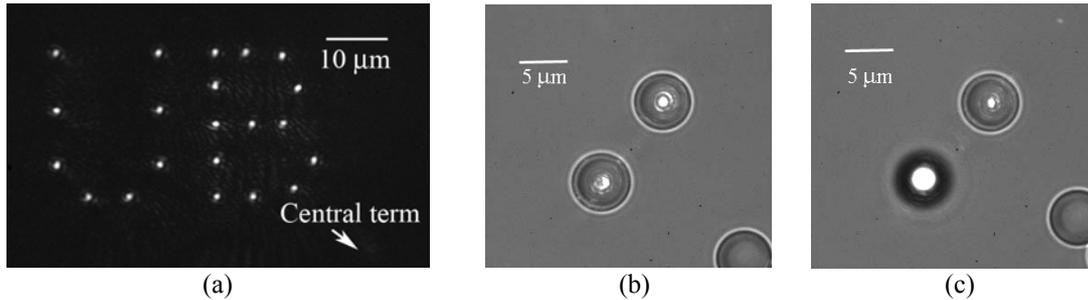

Figure 15. Experimental results with the Nikon microscope. (a) Several holographic optical traps, (b) Polystyrene microspheres (d=5 μm) trapped at the same plane or (c) at different depths.

## Acknowledgments

This research was funded by the Spanish Ministry of Education and Science through grants FIS2004-03450 and NAN2004-09348-C04-03. We are indebted to S. Vallmitjana for his general help and advice on several experimental issues.